\documentclass[conference,letterpaper]{IEEEtran}

\usepackage[utf8]{inputenc} 
\usepackage[T1]{fontenc}
\usepackage{url}
\usepackage{ifthen}
\usepackage{cite}
\usepackage[cmex10]{amsmath} 

\usepackage{authblk}

\RequirePackage{amsthm,amsfonts,amssymb, enumitem}
\usepackage{color}
\usepackage{bbm}
\usepackage{comment}

\makeatletter
\renewcommand*\env@matrix[1][*\c@MaxMatrixCols c]{%
  \hskip -\arraycolsep
  \let\@ifnextchar\new@ifnextchar
  \array{#1}}
\makeatother

\newcommand{\supp}{\mathrm{supp}\,}

\newcommand{\toas}{\overset{\text{a.s.}}{\to}}

\newcommand{\tr}{\mathrm{tr}\,}

\newcommand{\bV}{\mathbf{V}}
\newcommand{\bSigma}{\mathbf{\Sigma}}
\newcommand{\bx}{\mathbf{x}}

\newcommand{\bI}{\mathbf{I}}

\newcommand{\bW}{\mathbf{W}}

\newcommand{\bU}{\mathbf{U}}

\newcommand{\E}{\mathbb{E}}

\newcommand\diag{\mathrm{diag}}
\newcommand{\im}{\mathrm{i}\hspace{1.5pt}}

\newcommand{\bOne}{\mathbf{1}}

\newcommand{\NN}{\mathbb{N}}
\newcommand{\RR}{\mathbb{R}}
\newcommand{\CC}{\mathbb{C}}

\newcommand{\bA}{\boldsymbol{A}}

\newcommand{\dd}{\mathrm{d}}

\newcommand{\WADD}{\mathsf{WADD} }

\newcommand{\ARL}{\mathsf{ARL} }

\newcommand{\esssup}{\mathrm{ess \hspace{1.5pt} sup \hspace{2pt}}}

\theoremstyle{plain}

\newtheorem{theorem}{Theorem}[section]
\newtheorem*{theorem*}{Theorem}

\newtheorem{proposition}[theorem]{Proposition}

\newtheorem{definition}[theorem]{Definition}

\newcommand{\bOmega}{\boldsymbol{\Omega}}

\theoremstyle{remark}

\newtheorem{remark}{Remark}

\newcommand{\bmu}{\boldsymbol{\mu}}

\newcommand{\Aspect}{{\sc [Aspect($\gamma$)]}}

\newcommand{\Spec}{{\sc [Spec($H$)]}}

\newcommand{\Asy}{{\sc [Asy($\gamma$,$H$)]}}

\newcommand{\Reg}{{\sc [Shrink($\delta$)]}}

\usepackage{lipsum}

\newcommand\blfootnote[1]{%
  \begingroup
  \renewcommand\thefootnote{}\footnote{#1}%
  \addtocounter{footnote}{-1}%
  \endgroup
}

\begin{document}
\title{High-Dimensional Sequential Change Detection}

\author[1]{Robert Malinas}
\author[2]{Dogyoon Song}
\author[3]{Benjamin D. Robinson}
\author[1]{Alfred O. Hero III}
\affil[1]{Department of Electrical \& Computer Engineering, University of Michigan, Ann Arbor, MI}
\affil[2]{Department of Statistics, University of California-Davis, Davis, CA}
\affil[3]{Air Force Office of Scientific Research, Arlington, VA}
\affil[ ]{E-mail: rmalinas@umich.edu,  dgsong@ucdavis.edu, benjamin.robinson.8@us.af.mil,  hero@umich.edu}

\maketitle

\begin{abstract}
We address the problem of detecting a change in the distribution of a high-dimensional multivariate normal time series. 
Assuming that the post-change parameters are unknown and estimated using a window of historical data, we extend the framework of quickest change detection (QCD) to the high-dimensional setting in which the number of variables increases proportionally with the size of the window used to estimate the post-change parameters. 
Our analysis reveals that an information theoretic quantity, which we call the Normalized High-Dimensional Kullback-Leibler divergence (NHDKL),
governs the high-dimensional asymptotic performance of QCD procedures.
Specifically, we show that the detection delay is asymptotically inversely proportional to the difference between the NHDKL of the true post-change versus pre-change distributions and the NHDKL of the true versus estimated post-change distributions. In cases of perfect estimation, where the latter NHDKL is zero, the delay is inversely proportional to the NHDKL between the post-change and pre-change distributions alone. Thus, our analysis is a direct generalization of the traditional fixed-dimension, large-sample asymptotic framework, where the standard KL divergence is asymptotically inversely proportional to detection delay. Finally, we identify parameter estimators that asymptotically minimize the NHDKL between the true versus estimated post-change distributions, resulting in a QCD method that is guaranteed to outperform standard approaches based on fixed-dimension asymptotics.
\end{abstract}
\blfootnote{This research was partially supported by grants from a Fellowship from the  US Dept of Energy ETI Consortium, under grant DE-NA0003921, the  Army Research Office, grant number W911NF2310343, and the National Science Foundation, grant number CCF-2246213.}
\section{Introduction}
Suppose we sequentially observe independent high-dimensional multivariate normal samples $\bx_1, \bx_2, \ldots \in \RR^p$, $p  \gg 1$, with unknown change point $\nu \in \NN \cup \{0, \infty\}$, such that
\begin{equation} \label{eq:data}
\bx_t \sim
\begin{cases}
\mathcal{N}(\boldsymbol{0}, \bI), & \qquad \forall t \leq \nu, \\
\mathcal{N}(\bmu, \bSigma), & \qquad \forall t > \nu,
\end{cases}
\end{equation}
where $\bmu \in \RR^p$,  $\bSigma \in \mathbb{S}_+^{p \times p}$ is symmetric positive definite, $(\bmu, \bSigma) \neq (\boldsymbol{0}, \bI),$ and $\mathcal{N}(\boldsymbol{\xi}, \boldsymbol{\Xi})$ denotes the multivariate normal distribution with mean $\boldsymbol{\xi} \in \RR^p$ and covariance $\boldsymbol{\Xi} \in \mathbb{S}_+^{p \times p}$.
The problem addressed in this paper is to detect the change in distribution as quickly as possible after it occurs, assuming that the post-change parameters $\bmu$ and $\bSigma$ are unknown\footnote{If the pre-change distribution is not standard normal but its parameters are known, then one can always reduce to \eqref{eq:data} without loss of generality.} and the number of variables $p$ is large. The objective is to minimize the change point detection delay while maintaining the false alarm rate below a specified threshold.

This problem is one instance of a broad class of problems known as quickest change detection (QCD).
Originating in industrial quality control, where the observations \((\bx_t)_{t \geq 1}\) represent measurements from a manufacturing system and the change point $\nu$ signifies the onset of a manufacturing error, QCD has since been applied to various domains, including distributed sensing \cite{tartakovsky2008asymptotically}, cognitive radio \cite{lai2008quickest}, and intrusion detection in computer systems \cite{thottan2003anomaly}.

Modern applications often involve a vast number of variables \( p \). 
In scenarios where the post-change parameters \( \boldsymbol{\mu} \) and \( \boldsymbol{\Sigma} \) are unknown, traditional QCD methods rely on estimating these parameters using \( n \) historical samples. These methods typically require that the sample size \( n \) be substantially larger than the number of variables \( p \), which becomes impractical in high-dimensional contexts.

This paper focuses on efficiently detecting distribution changes of the form \eqref{eq:data} when the number of samples \(n\) used to estimate the unknown post-change parameters scales asymptotically with the number of variables \(p\), as both \(p\) and \(n\) grow to infinity: specifically, we assume that $p/n$ approaches a fixed positive and finite real number. 
In this high-dimensional regime, universal properties emerge due to the concentration of measure phenomenon \cite{ledoux2001concentration, van2014probability, vershynin2018high}.

Notable examples include the Marchenko-Pastur theorem \cite{marchenko1967distribution} from random matrix theory, which asymptotically characterizes the empirical distribution of the eigenvalues of random Gram matrices, as well as concentration of random quadratic forms \cite{rudelson2013hw, adamczak2015hanson, hanson1971bound} and random dot products \cite{hoeffding1963probability, chernoff1952measure, azuma1967weighted}.
By leveraging these high-dimensional phenomena, in this paper we extend the standard QCD theory for \eqref{eq:data} to the high-dimensional regime.

Our main results are twofold.
We consider a standard QCD procedure \cite{page1954continuous} with plug-in estimates of $\bmu$, $\bSigma$, estimated from a window of historical data \cite{xie2023window}, within the widely studied asymptotic regime where the false alarm rate approaches zero \cite{lorden1971procedures}, where we assume $n, p \rightarrow \infty$ and $p/n \rightarrow \gamma \in (0, \infty).$
In this high-dimensional regime, we demonstrate that the asymptotic Kullback-Leibler (KL) divergence between the true post-change distribution and its estimated counterpart, when normalized by the dimension \( p \), governs the increase in detection delay resulting from imperfect parameter estimation. We introduce this metric as the Normalized High-Dimensional Kullback-Leibler Divergence (NHDKL). Furthermore, we identify an optimal estimator that minimizes the NHDKL between the true and estimated post-change distributions within a broad class of parameter estimators, including standard maximum likelihood estimators for the post-change parameters \(\boldsymbol{\mu}\) and \(\boldsymbol{\Sigma}\). This optimization leads to a concrete QCD procedure for \eqref{eq:data} that exhibits provably superior asymptotic performance in high-dimensional settings compared to traditional QCD methods designed for fixed dimension \( p \).

\section{Non-Bayesian Quickest Change Detection}
We adopt the non-Bayesian QCD framework introduced in \cite{lorden1971procedures}.
Define the Worst Average Detection Delay ($\WADD$) and Average Run Length ($\ARL$), respectively, as
\begin{align*}
\WADD(\tau) & = \sup \limits_{\nu \geq 1} \esssup \E_{\nu} \left [ (\tau - \nu)^+ \big\vert \hspace{2pt} \bx_1 , \ldots, \bx_{\nu}  \right ], \\
\ARL(\tau) & =  \E_\infty \left [ \tau \right ],
\end{align*}
where $\tau$ is any stopping time with respect to $(\bx_t)_{t \geq 1},$ $(x)^+ = \max \{x, 0\}$ for any $x \in \RR$, and $\E_\nu$ denotes expectation with change point $\nu$.
Intuitively, $\WADD$ represents the worst-case average delay in detecting a change, considering the most unfavorable change point and pre-change observations. Conversely, $\ARL$ measures the expected time until a false alarm occurs when no change has taken place.
We adopt the standard objective of minimizing $\WADD$ under a constraint on $\ARL$:
\begin{align} \label{eq:Opt}
\tag{Opt}
\begin{split}
\textrm{minimize} \quad & \WADD(\tau), \\
\textrm{subject to} \quad & \ARL(\tau) \geq e^b, 
\end{split}
\end{align}
where $b > 0$ controls the false-alarm tolerance and the optimization variable $\tau$ ranges over the set of stopping times with respect to $(\bx_t)_{t \geq 1}$ \cite{lorden1971procedures, moustakides1986optimal, ritov1990decision}.
We note that both Bayesian approaches \cite{shiryaev1963optimum}, which treat the change point $\nu$ as an integer-valued random variable, and other non-Bayesian methods \cite{pollak1985optimal}, which replace $\WADD$ with a less conservative objective function, have also been studied.
See the monographs \cite{poor2008quickest, tartakovsky2014sequential, xie2021sequential, veeravalli2014quickest} for a comprehensive overview of QCD methods.

Let $f(\bx; \boldsymbol{\xi}, \boldsymbol{\Xi})$ denote the probability density function (pdf) of $\mathcal{N}(\boldsymbol{\xi}, \boldsymbol{\Xi})$ evaluated at $\bx \in \RR^p$, where $\boldsymbol{\xi} \in \RR^p$ and $\boldsymbol{\Xi} \in \mathbb{S}_+^{p \times p}$ is symmetric positive definite, and let $f_0 \equiv f(\cdot ; \boldsymbol{0}, \bI)$ denote the standard multivariate normal pdf on $\RR^p$.
If the post-change parameters $\bmu$ and $\bSigma$ are known, then the (optimal) solution to \eqref{eq:Opt} is given by the CuSum stopping time $\tau_c$ with stopping rule
\begin{equation} \label{eq:tau}
\tau_c = \inf \{t \geq 1 : Y_{t} \geq b \},
\end{equation}
where the CuSum statistic $Y_t$ is updated recursively as
\begin{equation} \label{eq:Yt}
Y_{t} = \left (Y_{t-1} + \log \dfrac{f(\bx_t; \bmu, \bSigma)}{f_0(\bx_t)} \right )^+, \qquad \forall t \geq 1, 
\end{equation}
with the initialization $Y_0 = 0$ \cite{page1954continuous}.
Optimality of $\tau_c$ with respect to \eqref{eq:Opt} was first proven in \cite{moustakides1986optimal} and later reaffirmed in \cite{ritov1990decision}, where it was shown that $\tau_c$ is a Bayes rule with respect to a loss criterion derived from a sequential stochastic game.

The expected post-change log-likelihood ratio (LLR) $\log \frac{f(\bx_t; \bmu, \bSigma)}{f_0(\bx_t)}$ appearing in \eqref{eq:Yt} is given by the KL divergence from $f(\cdot; \bmu, \bSigma)$ to $f_0$ given by the integral
\begin{align} \label{eq:EpostLLR}
D(f(\cdot ; \bmu, \bSigma) \Vert f_0) & = \int f(\bx; \bmu, \bSigma) \log \dfrac{f(\bx; \bmu, \bSigma)}{f_0(\bx)} \dd \bx \nonumber \\
& =  \dfrac{1}{2} \left [ \Vert \bmu \Vert_2^2 - \log \vert \bSigma \vert + \tr \bSigma - p\right ],
\end{align}
which is known to characterize the asymptotic performance of the CuSum procedure in the limit as the false alarm rate $1/\ARL(\tau_c)$ approaches zero. 
Specifically, if $D( f(\cdot; \bmu, \bSigma) \Vert f_0) \in (0, \infty)$, then \cite[Lemma 1]{xu2021optimum} gives
\begin{equation}
\lim_{b \rightarrow \infty} \dfrac{  D( f(\cdot; \bmu, \bSigma) \Vert f_0) \cdot \WADD(\tau_c)}{b} = 1,
\label{eq:tradeoff}
\end{equation}
which implies that the worst average delay  of the CuSum stopping time $\WADD(\tau_c)$ asymptotically decays at rate
$1/D( f(\cdot; \bmu, \bSigma) \Vert f_0)$ relative to the stopping threshold.
The regime $b \rightarrow \infty$ is of primary interest because one typically wants a large $\ARL$ to avoid frequent false alarms.

\section{Large-Sample Asymptotic QCD} \label{sec:largesample}
In practice, the post-change parameters $\bmu$ and $\bSigma$ are unknown.
To address this, a common approach is to relax \eqref{eq:Opt} and to consider asymptotic optimality of stopping times as their false alarm rate approaches zero.
Specifically, consider a sequence of stopping times $(\tau_n)_{n \geq 1}$ with the aim of solving
\begin{align} \label{eq:AsyOpt}
\tag{AsyOpt}
\begin{split}
\textrm{minimize} \quad & \limsup_{n \rightarrow \infty} p \cdot b_n^{-1}\WADD(\tau_n), \\
\textrm{subject to} \quad & \liminf_{n \rightarrow \infty} b_n^{-1}\log \left ( \ARL(\tau_n) \right ) \geq 1, 
\end{split}
\end{align}
where $(b_n)_{n \geq 1}$ is a user-defined sequence of positive real numbers such that $\lim \limits_{n \rightarrow \infty} b_n = \infty$, and $\tau_n$ is a stopping time with respect to $(\bx_t)_{t \geq 1}$ for each $n \geq 1$.
This problem is a natural relaxation of \eqref{eq:Opt} that admits solutions when the post-change parameters are unknown. 
The constraint on $\liminf \limits_{n \rightarrow \infty} b_n^{-1}\log \left ( \ARL(\tau_n) \right )$ guarantees a false alarm rate asymptotically no worse than that of the CuSum procedure with threshold $b_n$, and the objective function $\limsup \limits_{n \rightarrow \infty} p \cdot b_n^{-1}\WADD(\tau_n)$ serves to minimize detection delay in this limit.
The scale factor of $p$ in the objective function has no consequences in the fixed-dimension setting but will be useful for the forthcoming high-dimensional analysis.

When the post-change parameters $\bmu$ and $\bSigma$ are known, the sequence of CuSum stopping times $(\tau_{c,n})_{n \geq 1}$ given by
\begin{equation} \label{eq:tau_n}
\tau_{c, n} = \inf \{t \geq 1 : Y_{t} \geq b_n \},
\end{equation}
solves \eqref{eq:AsyOpt} and achieves the global minimum objective value of $\limsup \limits_{n \rightarrow \infty} p \cdot b_n^{-1}\WADD(\tau_{c,n}) = p/D( f(\cdot; \bmu, \bSigma) \Vert f_0)$   \eqref{eq:tradeoff}.
When $\bmu$ and $\bSigma$ are unknown, the CuSum statistic \eqref{eq:Yt} can not be computed.
In this setting, a prevailing approach is to estimate $\bmu$ and $\bSigma$ using historical data and incorporate the estimates into a CuSum-like procedure.
These methods, known as a plug-in procedures, are comprehensively reviewed in \cite{lai1995sequential}.
Next, we discuss one such methodology that solves \eqref{eq:AsyOpt} in the large-sample setting where the number of variables $p$ is fixed with respect to $n$.

The Window-Limited CuSum procedure (WLCuSum), defined in \cite{xie2023window},
considers estimating $\bmu$ and $\bSigma$ from a sliding window of historical observations and directly substituting the estimated parameters into the the CuSum recursion \eqref{eq:Yt}. 
Without loss of generality, we take the window length to be $n$. 
At time $t \geq n+1,$ consider the matrix of past samples $\bW_{t-1, n} = [\bx_{t - n} \ \cdots \ \bx_{t-1}] \in \RR^{p \times n}$ used to estimate $\bmu$ and $\bSigma$,
\[
\bx_{1},  \bx_{2} , \ldots, \underbrace{ \bx_{t-n} , \ldots, \bx_{t-1}}_{\textrm{past training window $\bW_{t-1, n}$}}, \underbrace{\bx_{t}}_{\textrm{sample at time $t$}}, \quad \underbrace{\bx_{t+1}, \ldots}_{\textrm{not yet observed}}
\]
Let $\mathcal{M}_n: \RR^{p \times n} \rightarrow \RR^{p}$ and $\mathcal{S}_n: \RR^{p \times n} \rightarrow \mathbb{S}_+^{p \times p}$ be estimators of $\bmu$ and $\bSigma$, respectively, and set $\hat{\bmu}_{t, n} = \mathcal{M}_n(\bW_{t, n})$ and $\hat{\bSigma}_{t, n} = \mathcal{S}_n(\bW_{t, n})$ for each $t \geq n.$
Define the $n$-th WLCuSum stopping time $\hat{\tau}_{c, n}$ of $(\bx_t)_{t \ \geq 1}$ as
\begin{equation} \label{eq:tauhat}
\hat{\tau}_{c, n} = \inf \{t \geq n + 1 : \hat{Y}_{t, n} \geq b_n \},
\end{equation}
where $b_n > 0$ is a user-defined stopping threshold, and\begin{equation} \label{eq:Yhat}
\hat{Y}_{t, n} = \left (\hat{Y}_{t-1, n} + \log \dfrac{f(\bx_{t}; \hat{\bmu}_{t-1, n}, \hat{\bSigma}_{t-1, n})}{f_0(\bx_{t})} \right )^+,
\end{equation}
for all $t \geq n + 1$, with the initialization $\hat{Y}_{n, n} = 0$.
In \cite{xie2023window}, it is shown that $\hat{\tau}_{c, n}$ solves \eqref{eq:AsyOpt} under the assumptions that
\begin{enumerate}
\item $\mathcal{M}_n$ and $\mathcal{S}_n$ are maximum likelihood estimators (MLEs);
\item the window size $n$ satisfies $n = \mathit{o}(b_n)$ as $n \rightarrow \infty$. 
\end{enumerate}

\section{High-Dimensional Asymptotic QCD}

To extend the QCD framework \eqref{eq:AsyOpt} to high-dimensional settings, where the number of variables $p$ increases proportionally with the window size $n$, we begin by defining a sequence of detection problems\footnote{This assumption is common in high-dimensional random matrix theory, most famously in Marchenko Pastur theorem \cite{marchenko1967distribution} in the analysis of sample covariance matrices.} indexed by the window size \( n \):
\begin{description}
  \item[\Aspect] The window length $n$ and the number of variables $p_n$ in the $n$-th problem 
    follow the  proportional-growth limit $p_n/n \rightarrow \gamma 
 \in (0, \infty)$ as $n \rightarrow \infty$.
\end{description}
\begin{remark}
Under \Aspect, we denote $p = p_n$ to emphasize the dependence on $n$.
For objects that depend on both the time index $t$ and problem index $n$, we will subscript by the time index followed by the problem index, e.g., $\hat{\bSigma}_{t, n}$ denotes the covariance estimate at time $t $ in the $n$-th problem, computed using the window of data $\bW_{t, n} \in \mathbb{R}^{p_n \times n}$.
For objects that depend on the problem index but not the time index, we use a single subscript $n$, e.g., $\bSigma_n$ denotes the population covariance matrix in the $n$-th problem.
We omit the subscript $n$ from the normal pdfs $f$ and $f_0$.
\end{remark}
Next, we define a loss function for analysis under \Aspect.
\begin{definition}[Loss function] \label{def:loss}
For each $n \geq 1$, define the loss function in the $n$-th problem as the normalized excess delay incurred by $\hat{\tau}_{c,n}$ relative to the optimum $\tau_{c,n}$,
\[
\mathcal{L}_n(\mathcal{M}_n, \mathcal{S}_n) = p_n \cdot b_n^{-1} [\WADD(\hat{\tau}_{c,n}) - \WADD(\tau_{c,n})].
\]
\end{definition}
The limit of $\mathcal{L}_n(\mathcal{M}_n, \mathcal{S}_n)$ as $n \rightarrow \infty$ under \Aspect, should it exist, characterizes the asymptotic loss in performance due to high-dimensional effects.
Next, we define the sample mean and sample covariance estimators which are the MLEs of $\bmu_n$ and $\bSigma_n$, respectively, given a sample of i.i.d. copies of $\bx_n \sim \mathcal{N}(\bmu_n, \bSigma_n)$.
\begin{definition}[Sample estimators] \label{def:samp}
For every $n \geq 1$ and data matrix $\bW_n \in \mathbb{R}^{p_n \times n}$, define the sample mean and sample covariance estimators, respectively, as
\[
\overline{\mathcal{M}}_n(\bW_n) = n^{-1} \bW_n \bOne_{n},
\]
and
\[
\overline{\mathcal{S}}_n(\bW_n) = n^{-1}(\bW_n - \overline{\mathcal{M}}_n(\bW_n) \bOne_{n}^\top)(\bW_n - \overline{\mathcal{M}}_n(\bW_n) \bOne_{n}^\top)^\top ,
\]
where $\bOne_{n} = (1,1, \ldots 1) \in \RR^{n}$.
\end{definition}
The main result of \cite{xie2023window} is that $\lim \limits_{n \rightarrow \infty} \mathcal{L}_n(\overline{\mathcal{M}}_n, \overline{\mathcal{S}}_n) = 0$ when $n = \mathit{o}(b_n)$ and the number of variables $p$ is fixed.
However, in the high-dimensional regime \Aspect, both the sample mean and sample covariance estimators are inconsistent estimators, hence, $\liminf \limits_{n \rightarrow \infty} \mathcal{L}_n(\overline{\mathcal{M}}_n, \overline{\mathcal{S}}_n) > 0.$
Moreover, when $p_n > n$, the sample covariance matrix is singular, rendering the WLCuSum statistic \eqref{eq:Yhat} with $\hat{\bSigma}_{t,n} = \overline{\mathcal{S}}_n(\bW_{t,n})$  undefined.
As a result of these inconsistencies and potential undefinedness, procedures that rely on consistent MLEs fail to solve the asymptotic optimization problem \eqref{eq:AsyOpt} in high-dimensional settings.
Thus, the standard finite-dimensional analysis of \eqref{eq:tauhat}-\eqref{eq:Yhat} for classical QCD, which relies on large-sample consistency of MLEs, is not applicable to high-dimensional analysis under \Aspect.

Our main contribution is a framework for evaluating the performance of the WLCuSum stopping time \eqref{eq:tauhat} in the  high-dimensional regime \Aspect.
After presenting this framework, we propose a method for high-dimensional non-Bayesian QCD that is guaranteed to asymptotically outperform the standard approach.

\subsection{High-Dimensional Asymptotic Optimization Framework}
For a diagonalizable matrix $\bA \in \RR^{p \times p}$, denote by $\lambda_1(\bA) \geq \lambda_2(\bA) \geq \cdots \geq \lambda_{p}(\bA)$ the eigenvalues of $\bA.$
Moreover, define the empirical spectral distribution function (ESDF) of $\bA$ as the cumulative distribution function (CDF)
\[
F_{\bA}(x) = \dfrac{1}{p} \sum_{k = 1}^p \mathbbm{1}\{\lambda_k(\bA) \leq x\}, \qquad \forall x \in \RR.
\]
For a proper CDF $F$, we define $\supp F$ as the support of the probability measure induced by $F$.
We adopt the following asymptotic population covariance model:
\begin{description}
  \item[\Spec] The ESDF $F_{\bSigma_n}$  of $\bSigma_n$ converges to a limiting proper CDF $H$ at every continuity point of $H$.
There exist $h_1, h_2 \in (0, \infty)$ such that $h_1 \leq h_2$ and 
\( \mathrm{supp}(H) \subset [h_1, h_2]\).
\end{description}
The \Spec \ covariance model, introduced in the pioneering work \cite{marchenko1967distribution}, is common in random matrix theory, and has been broadly applied across fields such as mathematical finance \cite{ledoit2017nonlinear}, adaptive array processing \cite{mestre2006finite}, wireless communications \cite{tulino2005impact}, ridge regression \cite{richards2021asymptotics},  deep learning \cite{pastur2023random}, and more.
A notable special case of \Spec \ is the `spiked' covariance model in principal components analysis \cite{johnstone2001distribution}, for which $H$ is a shifted Heaviside step function.
Change point detection for a special case of \eqref{eq:data} was studied under the spiked model in \cite{xie2020sequential}.
We often make the following composite assumption.
\begin{description}
\item[\Asy]Assume that \Aspect \ and \Spec \ hold for some $\gamma \in (0, \infty) \setminus \{1\}$ and proper CDF $H$ supported on a finite union of intervals.
Moreover, suppose there exist $B,D_\infty \in (0, \infty)$ such that $
\limsup \limits_{n \rightarrow \infty} \max\{ \Vert \bSigma_n \Vert, \ \Vert \bSigma_n^{-1} \Vert \} \leq B$ and $
\lim \limits_{n \rightarrow \infty} p_n^{-1}D(f(\cdot ; \bmu_n, \bSigma_n) \Vert f_0) \rightarrow D_\infty.$
\end{description}

We consider the class of covariance shrinkage estimators, defined below.
\begin{definition}[Shrinkage estimator] \label{def:shrinkage}
A covariance estimator $\mathcal{S}_n$ is called a shrinkage estimator with shrinkage function $\delta_n:[0, \infty) \rightarrow (0, \infty)$ if
\begin{enumerate}
\item $\mathcal{S}_n \overline{\mathcal{S}}_n = \overline{\mathcal{S}}_n \mathcal{S}_n$;
\item $\lambda_k(\mathcal{S}_n) = \delta_n(\lambda_k(\overline{\mathcal{S}}_n))$ for all $k \in [p_n].$
\end{enumerate}
\end{definition}
Pioneered by Stein \cite{stein1975estimation, stein1986lectures}, covariance shrinkage estimation is often employed in multivariate analysis for improving the condition number of the sample covariance matrix \cite{anderson1963asymptotic, chen2010shrinkage}, especially in the high-dimensional regime \Aspect \ where the sample covariance matrix is known to be ill-conditioned or singular \cite{ledoit2011eigenvectors, ledoit2020analytical, ledoit2018optimal, ledoit2022quadratic, donoho2018optimal, ledoit2004well, ledoit2017nonlinear}.

The following notations will be convenient for the forthcoming discussion.
Let $(\bW_n)_{n \geq 1}$ be a sequence of data matrices such that the columns of $\bW_n \in \mathbb{R}^{p_n \times n}$ are i.i.d. copies of $\bx_n \sim \mathcal{N}(\bmu_n, \bSigma_n),$ and set 
\[
\hat{\bmu}_n = \overline{\mathcal{M}}_n(\bW_n) \quad \textrm{and} \quad \hat{\bSigma}_n = \overline{\mathcal{S}}_n(\bW_n).
\]
Under \Aspect \ and \Spec, the Marchenko-Pastur theorem \cite{marchenko1967distribution} dictates that
\[
F_{\overline{\mathcal{S}}_n(\bW_n)}(x) \toas G(x),
\]
for every continuity point $x$ of $G$, where $G$ is a deterministic proper CDF called the limiting spectral distribution function (LSDF) of the sequence of random matrices $(\overline{\mathcal{S}}_n(\bW_n))_{n \geq 1}.$
Moreover, $\supp G$ is compact.
Define the Stieltjes transform of $G$,
\[
m_G(z) = \int \dfrac{1}{x - z} G(\dd x), \qquad \forall z \in \CC_+,
\]
and its real-line limit
\[
\check{m}_G(x) = \lim_{\eta \rightarrow 0} m_G(x + \im \eta), \qquad \forall x \in \RR \setminus \{0\},
\]
the latter of which is shown to exist in \cite{silverstein1995analysis}.

We make the following assumption on the sequence of covariance estimators $(\mathcal{S}_n)_{n \geq 1}$:
\begin{description}
\item[\Reg] For each $n \geq 1$, the covariance estimator $\mathcal{S}_n$ is a shrinkage estimator with shrinkage function $\delta_n$. There exists a nonrandom and continuously differentiable limiting shrinkage function $\delta: [0, \infty) \rightarrow (0, \infty)$ such that $\delta_n \toas \delta$ uniformly on a compact neighborhood\footnote{We adopt the convention that a compact neighborhood of a point (resp. set) is any compact set that contains an open neighborhood of the point (resp. set).} of $G$.
\end{description}

\begin{theorem} \label{thm:main}
Assume \Asy, and suppose that \Reg \ holds for some limiting shrinkage function $\delta$.
Then the sequence of WLCuSum stopping times $(\hat{\tau}_{c, n})_{n \geq 1}$ with corresponding stopping thresholds $(b_n)_{n \geq 1}$ satisfies
\begin{enumerate}
\item $\ARL(\hat{\tau}_{c,n}) \geq e^{b_n}$ for every $n \geq 1$;
\vspace{3pt}
\item $\WADD(\hat{\tau}_{c,n}) = \E_0[\hat{\tau}_{c,n}]$ for every $n \geq 1$.
\end{enumerate}
Moreover, if  $(\mathcal{M}_n)_{n \geq 1} = (\overline{\mathcal{M}}_n)_{n \geq 1}$ and $n = \mathit{o} \big ( \sqrt{b_n}  \big ) \quad \textrm{as} \quad n \rightarrow \infty$, then
\[
\lim_{n \rightarrow \infty} \mathcal{L}_n(\overline{\mathcal{M}}_n, \mathcal{S}_n) \toas \mathcal{L}_\infty(\delta),
\]
where
\[
\mathcal{L}_\infty(\delta) = \dfrac{ \mathcal{D}_\infty(\delta)}{D_\infty[D_\infty - \mathcal{D}_\infty(\delta)]},
\]
and
\begin{align*}
\mathcal{D}_\infty(\delta) = \frac{1}{2} \Bigg [ \int & \dfrac{x}{\vert 1 - \gamma  - \gamma \check{m}_G(x)\vert^2 \delta(x) + \log(\delta(x))} G(\dd x) \\
& - \int \log(y) H(\dd y) - 1 \Bigg ].
\end{align*}
Furthermore,
\[
\mathcal{D}_\infty(\delta) = \lim_{n \rightarrow \infty} p_n^{-1} D( f(\cdot; \bmu, \bSigma) \Vert f(\cdot; \hat{\bmu}_n, \hat{\bSigma}_n)),
\]
almost surely.
\end{theorem}

When the parameters are known, the asymptotic tradeoff between the worst average detection delay, $\WADD(\tau_{c,n})$, and the average run length, $\ARL(\tau_{c,n})$, of the CuSum procedure is characterized in \eqref{eq:tradeoff} by the KL divergence $D(f(\cdot ; \bmu_n, \bSigma_n) \Vert f_0)$, as defined in \eqref{eq:EpostLLR}.
In contrast, when the parameters are estimated in the high-dimensional regime \Aspect, there is a loss incurred due to imperfect parameter estimation.
Asymptotically, this loss is captured by the quantity $\mathcal{D}_\infty(\delta),$ which we call the normalized high-dimensional KL divergence (NHDKL) between the true post-change distribution and estimated post-change distribution.
Therefore, the loss in performance exhibited in high dimensions is captured by an intuitive information-theoretic loss function.
Similarly, we call $D_\infty$ the NHDKL between the post-change distribution and pre-change distribution.

Fixing the sample mean estimator and sequence of shrinkage functions $(\delta_n)_{n \geq 1}$ as in Theorem \ref{thm:main}, we have
\[
 p_n b_n^{-1} \WADD(\hat{\tau}_{c,n}) = \mathcal{L}_n(\overline{\mathcal{M}}_n, \mathcal{S}_n) + p_n b_n^{-1} \WADD(\tau_{c,n})
\]
by the definition of $\mathcal{L}_n(\overline{\mathcal{M}}_n, \mathcal{S}_n)$.
By Theorem \ref{thm:main}, we obtain
\[
\lim_{n \rightarrow \infty} p_n b_n^{-1} \WADD(\hat{\tau}_{c,n}) \toas \mathcal{L}_\infty(\delta) + D_\infty^{-1},
\]
where we used the fact that $\lim \limits_{n \rightarrow \infty} p_n b_n^{-1} \WADD(\tau_{c,n}) = D_\infty^{-1}$ by a straightforward extension of \eqref{eq:tradeoff}.
By Theorem \ref{thm:main}, this gives
\begin{align*}
 p_n b_n^{-1} \WADD(\hat{\tau}_{c,n}) & \toas  \dfrac{ \mathcal{D}_\infty(\delta)}{D_\infty[D_\infty - \mathcal{D}_\infty(\delta)]} + \dfrac{1}{D_\infty} \\
& = [D_\infty - \mathcal{D}_\infty(\delta)]^{-1}.
\end{align*}
Hence, the asymptotic $\WADD$ of $\hat{\tau}_{c,n}$ is inversely proportional to the difference of the NHDKL of the post-change versus pre-change distributions and the NHDKL of the true vs estimated post-change distributions, directly extending \eqref{eq:tradeoff} to the case where the parameters are unknown in the limit \Aspect.
The large-sample limit  with fixed dimension $p$ can be recovered from this analysis by considering $\gamma \rightarrow 0.$
In this case, the sample mean $\overline{\mathcal{M}}_n$ and sample covariance $\overline{\mathcal{S}}_n$, the latter of which is a shrinkage estimator with $\delta_n(x) = x$ for all $x \in [0, \infty)$ and all $n \geq p$, are consistent with $p$ fixed as $n \rightarrow \infty$.
Hence, $\mathcal{D}(\delta) = 0$ and $\lim \limits_{n \rightarrow \infty} p b_n^{-1} \WADD(\hat{\tau}_{c,n}) \toas p/D_\infty,$ recovering the optimal tradeoff \eqref{eq:tradeoff}.

\subsection{High-Dimensional Asymptotically Optimal QCD Procedure}
The rest of the paper is devoted to deriving an asymptotically optimal shrinkage estimator that, together with the sample mean estimator, achieves asymptotically minimum detection delay among all shrinkage estimators in the setting of Theorem \ref{thm:main}.
The following proposition demonstrates that the minimizer of $\mathcal{D}_\infty$ coincides with that of $\mathcal{L}_\infty$.
\begin{proposition}
Let $\delta_1$ and $\delta_2$ be limiting shrinkage functions that satisfy {\sc Shrink($\delta_1$)} and {\sc Shrink($\delta_2$)}, respectively.
If
$\mathcal{D}_\infty(\delta_1) \leq \mathcal{D}_\infty(\delta_2),$ then
$\mathcal{L}_\infty(\delta_1) \leq \mathcal{L}_\infty(\delta_2).$
\end{proposition}

Therefore, to optimize performance of $\hat{\tau}_{c,n},$ it is sufficient to consider optimization over the set of limiting shrinkage functions $C^1 \big ( [0, \infty), (0, \infty) \big )$.
Under the assumptions of Theorem \ref{thm:main}, $\mathcal{D}_\infty(\delta)$ is the limit of the inverse Stein's loss, defined next, between the shrinkage estimate $\hat{\bSigma}_n$ and the true covariance matrix $\bSigma_n$.
\begin{definition} \label{def:ISLoss}
Define the inverse Stein's loss
\[
\mathfrak{L}_n^{IS}(\bA, \bSigma_n) = p_n^{-1} \tr(\bSigma_n \bA^{-1}) - p_n^{-1} \log \left \vert \bSigma_n\bA^{-1} \right \vert - 1, 
\]
for all $\bA \in \mathbb{S}_+^{p_n \times p_n}$, for all $n \geq 1.$
\end{definition}
\begin{proposition}
Under the assumptions of Theorem \ref{thm:main},
\[
\frac{1}{2} \mathfrak{L}_n^{IS}(\hat{\bSigma}_n; \bSigma_n) \toas \mathcal{D}_\infty(\delta) \quad \textrm{as} \quad n \rightarrow \infty.
\]

\end{proposition}

The following covariance estimator, derived in \cite{ledoit2022quadratic}, is a finite-dimensional approximation to the minimizer of $\mathcal{D}_\infty$ over $C^1 \big ( [0, \infty), (0, \infty) \big )$.

\begin{definition}[Ledoit-Wolf quadratic inverse Stein's loss shrinkage estimator (LWISE)] \label{def:LW}
For any $\bW \in \RR^{p \times n},$ write the singular value decomposition
\[
\bW - n^{-1} \bW \bOne_n \bOne_n^\top = \bU \bOmega \bV^\top,
\]
where $\bOmega = \diag(\omega_1, \ldots, \omega_{\min\{p, n\}}) \in \RR^{p \times n} $, and let $\gamma_n = p/(n-1).$
For $p \neq n - 1$, the Ledoit-Wolf quadratic inverse Stein's loss shrinkage function $\delta_n$ is defined as
\begin{align*}
& \delta^*_n(\omega)  = \\
& \begin{cases}
  \omega^{2} \big [ ( 1 - \gamma_n )^2 + 2 \gamma_n ( 1 - \gamma_n  ) g(\omega) + \gamma_n^2 h(\omega^{-1}) \big ]^{-1}, & p < n -1, \\
 \omega^2[h(\omega^{-1})]^{-1}, & p > n - 1,
\end{cases}
\end{align*}
for all $\omega > 0,$ where $\gamma_n = p/n$,
\begin{align*}
& g(x) = \min\{p,n\}^{-1} \sum_{i = 1}^{\min\{p,n\}} \omega_i^{-2} \dfrac{\omega_i^{-2} - x}{(\omega_i^{-2} - x)^2 + n^{-2/3} \omega_i^{-4}},
\end{align*}
and
\begin{align*}
& h(x) = g^2(x) \\
& +  \left [ \min\{p,n\}^{-1} \sum_{i = 1}^{\min\{p,n\}} \omega_i^{-2} \dfrac{n^{-1/3}\omega_i^{-2}}{(\omega_i^{-2} - x)^2 + n^{-2/3} \omega_i^{-4}} \right ]^2,
\end{align*}
for all $x \geq 0.$
Finally,
\[
\delta^*_n(0) = \bigg [  (\gamma_n - 1) \min \{p, n\}^{-1} \sum_{i = 1}^{\min \{p, n\}} \omega_i^{-2} \bigg ]^{-1}.
\]
\end{definition}

The following practical theorem states that, with the sample mean estimator fixed, LWISE achieves asymptotically lower $\WADD$ than any other shrinkage estimator.
Notably, the sample covariance matrix is a shrinkage estimator, hence LWISE achieves superior high-dimensional performance compared to the the classical MLE.
Therefore, we recommend using the sample mean estimator and LWISE covariance estimator for quickest change detection in the high-dimensional regime \Aspect.
\begin{theorem}
Assume \Asy.
Then $\delta_n^* \toas \delta^*$ uniformly on a compact neighborhood of $\supp G$, where $\delta^*$ is nonrandom and continuously differentiable.
Moreover, if $\mathcal{M}_n = \overline{\mathcal{M}}_n$, then
\[
\mathcal{L}_\infty(\delta^*) \leq \mathcal{L}_\infty(\delta), \qquad \forall \delta \in C^1([0, \infty), (0, \infty)),
\]
hence, LWISE achieves a lower value of $\limsup \limits_{n \rightarrow \infty} p_n \cdot b_n^{-1}\WADD(\hat{\tau}_{c,n})$ than any shrinkage estimator satisfying \Reg.
\end{theorem}

\section{Conclusion}
We addressed the detection of changes in the distribution of a high-dimensional multivariate normal time series by extending classical quickest change detection to scenarios where the number of variables scales with the estimation window size. We introduced the Normalized High-Dimensional Kullback-Leibler divergence (NHDKL) as a key metric governing the asymptotic performance of QCD procedures. Additionally, we identified parameter estimators that minimized this divergence, ensuring superior performance compared to standard approaches derived using fixed-dimension asymptotics.


\bibliographystyle{plain}
\bibliography{msrhISIT2025}

\begin{thebibliography}{10}

\bibitem{adamczak2015hanson}
Radoslaw Adamczak.
\newblock {A note on the Hanson-Wright inequality for random vectors with dependencies}.
\newblock {\em Electronic Communications in Probability}, 20(none):1 -- 13, 2015.

\bibitem{anderson1963asymptotic}
Theodore~Wilbur Anderson.
\newblock Asymptotic theory for principal component analysis.
\newblock {\em The Annals of Mathematical Statistics}, 34(1):122--148, 1963.

\bibitem{azuma1967weighted}
Kazuoki Azuma.
\newblock {Weighted sums of certain dependent random variables}.
\newblock {\em Tohoku Mathematical Journal}, 19(3):357 -- 367, 1967.

\bibitem{chen2010shrinkage}
Yilun Chen, Ami Wiesel, Yonina~C Eldar, and Alfred~O Hero.
\newblock Shrinkage algorithms for mmse covariance estimation.
\newblock {\em IEEE transactions on signal processing}, 58(10):5016--5029, 2010.

\bibitem{chernoff1952measure}
Herman Chernoff.
\newblock {A Measure of Asymptotic Efficiency for Tests of a Hypothesis Based on the sum of Observations}.
\newblock {\em The Annals of Mathematical Statistics}, 23(4):493 -- 507, 1952.

\bibitem{donoho2018optimal}
David~L. Donoho, Matan Gavish, and Iain~M. Johnstone.
\newblock Optimal shrinkage of eigenvalues in the spiked covariance model.
\newblock {\em Annals of Statistics}, 46(4):1742, 2018.

\bibitem{hanson1971bound}
F.~T. Hanson, D. L.;~Wright.
\newblock A bound on tail probabilities for quadratic forms in independent random variables.
\newblock {\em The Annals of mathematical statistics}, 42(3):1079–1083, 1971.

\bibitem{hoeffding1963probability}
Wassily Hoeffding.
\newblock Probability inequalities for sums of bounded random variables.
\newblock {\em Journal of the American Statistical Association}, 58(301):13--30, 1963.

\bibitem{johnstone2001distribution}
Iain~M. Johnstone.
\newblock {On the distribution of the largest eigenvalue in principal components analysis}.
\newblock {\em The Annals of Statistics}, 29(2):295 -- 327, 2001.

\bibitem{lai2008quickest}
Lifeng Lai, Yijia Fan, and H~Vincent Poor.
\newblock Quickest detection in cognitive radio: A sequential change detection framework.
\newblock In {\em IEEE GLOBECOM 2008-2008 IEEE Global Telecommunications Conference}, pages 1--5. IEEE, 2008.

\bibitem{lai1995sequential}
Tze~Leung Lai.
\newblock Sequential changepoint detection in quality control and dynamical systems.
\newblock {\em Journal of the Royal Statistical Society. Series B (Methodological)}, 57(4):613--658, 1995.

\bibitem{ledoit2011eigenvectors}
Olivier Ledoit and Sandrine P{\'e}ch{\'e}.
\newblock Eigenvectors of some large sample covariance matrix ensembles.
\newblock {\em Probability Theory and Related Fields}, 151(1):233--264, Oct 2011.

\bibitem{ledoit2004well}
Olivier Ledoit and Michael Wolf.
\newblock A well-conditioned estimator for large-dimensional covariance matrices.
\newblock {\em Journal of multivariate analysis}, 88(2):365--411, 2004.

\bibitem{ledoit2017nonlinear}
Olivier Ledoit and Michael Wolf.
\newblock Nonlinear shrinkage of the covariance matrix for portfolio selection: Markowitz meets goldilocks.
\newblock {\em The Review of Financial Studies}, 30(12):4349--4388, 2017.

\bibitem{ledoit2018optimal}
Olivier Ledoit and Michael Wolf.
\newblock {Optimal estimation of a large-dimensional covariance matrix under Stein’s loss}.
\newblock {\em Bernoulli}, 24(4B):3791 -- 3832, 2018.

\bibitem{ledoit2020analytical}
Olivier Ledoit and Michael Wolf.
\newblock {Analytical nonlinear shrinkage of large-dimensional covariance matrices}.
\newblock {\em The Annals of Statistics}, 48(5):3043 -- 3065, 2020.

\bibitem{ledoit2022quadratic}
Olivier Ledoit and Michael Wolf.
\newblock {Quadratic shrinkage for large covariance matrices}.
\newblock {\em Bernoulli}, 28(3):1519 -- 1547, 2022.

\bibitem{ledoux2001concentration}
Michel Ledoux.
\newblock {\em The concentration of measure phenomenon}.
\newblock Number~89. American Mathematical Soc., 2001.

\bibitem{lorden1971procedures}
Gary Lorden.
\newblock {Procedures for Reacting to a Change in Distribution}.
\newblock {\em The Annals of Mathematical Statistics}, 42(6):1897 -- 1908, 1971.

\bibitem{marchenko1967distribution}
Volodymyr Mar{\v{c}}enko and Leonid Pastur.
\newblock Distribution of eigenvalues for some sets of random matrices.
\newblock {\em Mathematics of the {USSR}-Sbornik}, 1(4):457--483, apr 1967.

\bibitem{mestre2006finite}
Xavier Mestre and Miguel~A. Lagunas.
\newblock Finite sample size effect on minimum variance beamformers: optimum diagonal loading factor for large arrays.
\newblock {\em IEEE Transactions on Signal Processing}, 54(1):69--82, 2006.

\bibitem{moustakides1986optimal}
George~V. Moustakides.
\newblock {Optimal Stopping Times for Detecting Changes in Distributions}.
\newblock {\em The Annals of Statistics}, 14(4):1379 -- 1387, 1986.

\bibitem{page1954continuous}
E.~S. Page.
\newblock {Continuous Inspection Schemes}.
\newblock {\em Biometrika}, 41(1-2):100--115, 06 1954.

\bibitem{pastur2023random}
Leonid Pastur and Victor Slavin.
\newblock On random matrices arising in deep neural networks: General i.i.d. case.
\newblock {\em Random Matrices: Theory and Applications}, 12(01):2250046, 2023.

\bibitem{pollak1985optimal}
Moshe Pollak.
\newblock Optimal detection of a change in distribution, 1985.

\bibitem{poor2008quickest}
{H. Vincent} Poor and Olympia Hadjiliadis.
\newblock {\em Quickest detection}, volume 9780521621045.
\newblock Cambridge University Press, United Kingdom, January 2008.
\newblock Publisher Copyright: {\textcopyright} Cambridge University Press 2009.

\bibitem{richards2021asymptotics}
Dominic Richards, Jaouad Mourtada, and Lorenzo Rosasco.
\newblock Asymptotics of ridge (less) regression under general source condition.
\newblock In {\em International Conference on Artificial Intelligence and Statistics}, pages 3889--3897. PMLR, 2021.

\bibitem{ritov1990decision}
Ya'acov Ritov.
\newblock Decision theoretic optimality of the cusum procedure.
\newblock {\em The Annals of Statistics}, 18(3):1464--1469, 1990.

\bibitem{rudelson2013hw}
Mark Rudelson and Roman Vershynin.
\newblock {Hanson-Wright inequality and sub-gaussian concentration}.
\newblock {\em Electronic Communications in Probability}, 18(none):1 -- 9, 2013.

\bibitem{shiryaev1963optimum}
Albert~N. Shiryaev.
\newblock On optimum methods in quickest detection problems.
\newblock {\em Theory of Probability \& Its Applications}, 8(1):22--46, 1963.

\bibitem{silverstein1995analysis}
Jack~W Silverstein and Sang-Il Choi.
\newblock Analysis of the limiting spectral distribution of large dimensional random matrices.
\newblock {\em Journal of Multivariate Analysis}, 54(2):295--309, 1995.

\bibitem{stein1986lectures}
C~Stein.
\newblock Lectures on the theory of estimation of many parameters.
\newblock {\em Journal of Mathematical Sciences}, 34(1):1373--1403, 1986.

\bibitem{stein1975estimation}
Charles Stein.
\newblock Estimation of a covariance matrix.
\newblock In {\em 39th Annual Meeting IMS, Atlanta, GA, 1975}, 1975.

\bibitem{tartakovsky2014sequential}
Alexander Tartakovsky, Igor Nikiforov, and Michel Basseville.
\newblock {\em Sequential Analysis: Hypothesis Testing and Changepoint Detection}.
\newblock Chapman and Hall/CRC, 1st edition, 2014.

\bibitem{tartakovsky2008asymptotically}
Alexander~G Tartakovsky and Venugopal~V Veeravalli.
\newblock Asymptotically optimal quickest change detection in distributed sensor systems.
\newblock {\em Sequential Analysis}, 27(4):441--475, 2008.

\bibitem{thottan2003anomaly}
Marina Thottan and Chuanyi Ji.
\newblock Anomaly detection in ip networks.
\newblock {\em IEEE Transactions on signal processing}, 51(8):2191--2204, 2003.

\bibitem{tulino2005impact}
Antonia~M. Tulino, Angel Lozano, and Sergio Verdu.
\newblock Impact of antenna correlation on the capacity of multiantenna channels.
\newblock {\em IEEE Transactions on Information Theory}, 51(7):2491--2509, 2005.

\bibitem{van2014probability}
Ramon Van~Handel.
\newblock Probability in high dimension.
\newblock {\em Lecture Notes (Princeton University)}, 2(3):2--3, 2014.

\bibitem{veeravalli2014quickest}
Venugopal~V. Veeravalli and Taposh Banerjee.
\newblock Chapter 6 - quickest change detection.
\newblock In Abdelhak~M. Zoubir, Mats Viberg, Rama Chellappa, and Sergios Theodoridis, editors, {\em Academic Press Library in Signal Processing: Volume 3}, volume~3 of {\em Academic Press Library in Signal Processing}, pages 209--255. Elsevier, 2014.

\bibitem{vershynin2018high}
Roman Vershynin.
\newblock {\em High-Dimensional Probability: An Introduction with Applications in Data Science}, volume~47.
\newblock Cambridge University Press, 2018.

\bibitem{xie2023window}
Liyan Xie, George~V. Moustakides, and Yao Xie.
\newblock Window-limited cusum for sequential change detection.
\newblock {\em IEEE Transactions on Information Theory}, 69(9):5990--6005, 2023.

\bibitem{xie2020sequential}
Liyan Xie, Yao Xie, and George~V Moustakides.
\newblock Sequential subspace change point detection.
\newblock {\em Sequential Analysis}, 39(3):307--335, 2020.

\bibitem{xie2021sequential}
Liyan Xie, Shaofeng Zou, Yao Xie, and Venugopal~V. Veeravalli.
\newblock Sequential (quickest) change detection: Classical results and new directions.
\newblock {\em IEEE Journal on Selected Areas in Information Theory}, 2(2):494--514, 2021.

\bibitem{xu2021optimum}
Qunzhi Xu, Yajun Mei, and George~V. Moustakides.
\newblock Optimum multi-stream sequential change-point detection with sampling control.
\newblock {\em IEEE Transactions on Information Theory}, 67(11):7627--7636, 2021.

\end{thebibliography}

\end{document}